# Nanostructural Origin of Semiconductivity and Large Magnetoresistance in Epitaxial NiCo$_2$O$_4$/Al$_2$O$_3$ Thin Films

Congmian Zhen,[a,b] XiaoZhe Zhang,[b,c] Wengang Wei,[a,d] Wenzhe Guo,[a] Ankit Pant,[b] Xiaoshan Xu,[*,b] Jian Shen,[d] Li Ma,[a] and Denglu Hou[a]

[a]Hebei Advanced Thin Films Laboratory, Department of Physics, Hebei Normal University, Shijiazhuang 050024, China

[b]Department of Physics and Astronomy, Nebraska Center for Materials and Nanoscience, University of Nebraska-Lincoln, NE 68588, USA

[c]Department of Material Physics, School of Science, Xi'an jiaotong University, 710049, China

[d]State Key Laboratory of Surface Physics and Department of Physics, Fudan University, Shanghai 200433, China

**Abstract**

Despite low resistivity (~ 1 mΩ cm), metallic electrical transport has not been commonly observed in the inverse spinel NiCo$_2$O$_4$, except in certain epitaxial thin films. Previous studies have stressed the effect of valence mixing and degree of spinel inversion on the electrical conduction of NiCo$_2$O$_4$ films. In this work, we studied the effect of nanostructural disorder by comparing the NiCo$_2$O$_4$ epitaxial films grown on MgAl$_2$O$_4$ (111) and on Al$_2$O$_3$ (001) substrates. Although the optimal growth conditions are similar for the NiCo$_2$O$_4$ (111)/MgAl$_2$O$_4$ (111) and the NiCo$_2$O$_4$ (111)/Al$_2$O$_3$ (001) films, they show metallic and semiconducting electrical transport respectively. Post-growth annealing decreases the resistivity of NiCo$_2$O$_4$ (111)/Al$_2$O$_3$ (001) films, but the annealed films are still semiconducting. While the semiconductivity and the large magnetoresistance in NiCo$_2$O$_4$ (111)/Al$_2$O$_3$ (001) films cannot be accounted for in terms of the non-optimal valence mixing and spinel inversion, the presence of anti-phase boundaries between nano-sized crystallites, generated by the structural mismatch between $NiCo_2O_4$ and $Al_2O_3$, may explain all the experimental observations in this work. These results reveal nanostructural disorder as another key factor in controlling the electrical transport of $NiCo_2O_4$, with potentially large magnetoresistance for spintronics applications.

---

[*] Corresponding authors. Email: xiaoshan.xu@unl.edu; cmzhen@hebtu.edu.cn

## 1. Introduction

Recent discovery of metallicity in NiCo$_2$O$_4$ (NCO) has diversified the functional properties of the spinel material family in addition to their celebrated ferrimagnetism (e.g. in Fe$_3$O$_4$) [1-4]. The high conductivity and advantageous electrochemical properties of NCO is compelling for applications for electrode in energy storage devices such as metal-ion batteries and electrochemical supercapacitors [5-13]. On the other hand, the metallic conduction of NCO, i.e. low resistivity even at low temperature, has only been observed in epitaxial thin films prepared in certain conditions [1-4], while in most cases insulating (or semiconducting) behavior has been reported [4,14-21]. Therefore, many factors, including crystal structure, nanostructure, and electronic structure, are believed to be critical in the mechanism of electrical conduction in NCO.

NiCo$_2$O$_4$ has an inverse spinel crystal structure. In the unit cell, Co ions occupy sites with a tetrahedral (T$_d$) local environment, while Co ions and Ni ions share the sites with an octahedral (O$_h$) local environment, as illustrated in Figure 1 [2,22]. The magnetic moments of the Ni and Co ions on O$_h$ and T$_d$ sites respectively, are believed to be anti-aligned, corresponding to a ferrimagnetic order below $T_C \approx 330$ K [1,2]. In contrast, the Co$^{3+}$ ions on the O$_h$ sites do not contribute to the magnetization due to the zero-spin state (e$_g^0$ t$_{2g}^6$, S = 0) [22-25]. For the polycrystalline NCO, the measured resistivity always increases rapidly on cooling, corresponding to a semiconducting behavior [15-21]. In contrast, high conductivity at low temperature has been observed in epitaxial thin films grown on MgAl$_2$O$_4$ (MAO) and MgO substrates, indicative of metallicity [1-4]. The study of metallic and semiconducting NCO thin films, grown on the MAO substrates under different conditions, suggests that the mixed valence of Ni$^{2+}$ and Ni$^{3+}$ on the O$_h$ site and the double-exchange interactions are critical for the metallicity [2]. The comparison of Raman spectroscopy of metallic and semiconducting NCO/MAO thin films demonstrate that the cation disorder on the O$_h$ sites favors the metallicity [26]. Furthermore, the occupation of Ni on the O$_h$ site instead of the T$_d$ site was shown to be important for high conductivity in the textured NCO films grown on SrTiO$_3$ (STO) substrates [14].

In this work, we focus on the effect of nanostructure of NCO films on the conductivity. Transition between a metal and an insulating (or semiconducting) phase may be caused by changes of electronic structures, such as changes of band overlap and band filling; it may also be caused by the disorder which localizes electronic states [27,28]. As shown in **Figure** 1, the mixed valence (Ni$^{2+}$ and Ni$^{3+}$) on the O$_h$ sites, allows the hopping of localized state (polaron), similar to that of Fe (Fe$^{2+}$ and Fe$^{3+}$) on the O$_h$ sites in Fe$_3$O$_4$. As proposed previously, when the degree of valence mixing is high enough, the double-exchange interaction may make the e$_g$ states on Ni itinerant, generating metallic conduction [2]. Besides the degree of valence mixing, structural and nanostructural disorder may also be important in the localization of the electrons. To investigate the effect of disorder on the nanostructure, we have studied the epitaxial NCO thin films grown on both MAO and Al$_2$O$_3$ (ALO) substrates. Due to the significant difference between the crystal structures of NCO and ALO, structural antiphase boundary is expected to exist and play a role in the conductivity. We found that although the optimal growth condition between the NCO (111) /ALO (001) and NCO (111)/MAO (111) films are similar, the NCO (111)/ALO (001) films show a semiconducting behavior and a large dependence of resistivity on the film thickness. These results indicate the sensitivity of electrical transport of NCO on the nanostructures, which provide insight in understanding the loss of metallicity in most polycrystalline samples.

## 2. Experimental Section

Pulsed laser deposition (PLD) was employed to grow epitaxial NCO (111) thin films on 5 mm × 5 mm ALO (001) and MAO (111) substrates, with various O$_2$ pressure (5 to 50 mTorr), growth temperature (300 to 500 ˚C), and thickness (24 to 95 nm) with a KrF excimer laser (λ= 248 nm, frequency = 10 Hz, fluence = 2.5 mJ/cm$^2$). Epitaxial relation between the film and substrate and the surface morphology of the films, are monitored in-situ by the reflection high energy electron diffraction (RHEED). The crystallinity, thickness, and out-of-plane lattice constants of the films were measured using x-ray diffraction (XRD) with a Rigaku D/Max-B x-ray diffractometer (Co K-α radiation, λ= 1.789 Å) and a Rigaku SmartLab x-ray diffractometer (copper K-α source, λ = 1.54 Å).

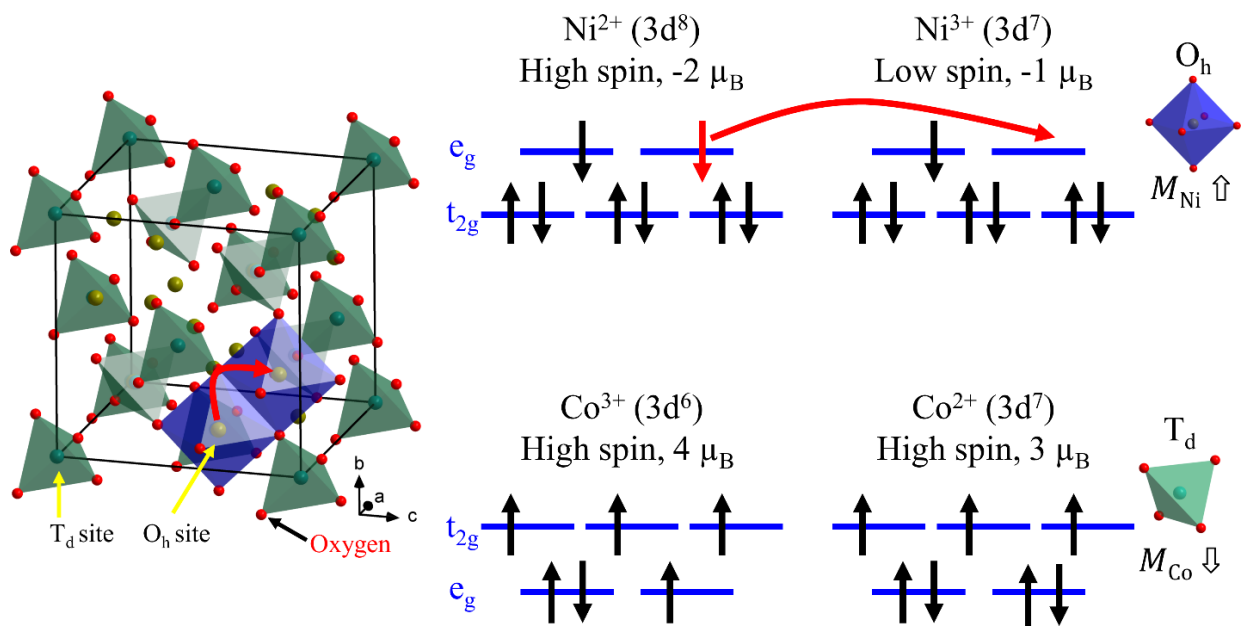

**Figure 1**. Left: Atomic model of the unit cell of $NiCo_2O_4$ in the inverse spinel crystal structure. The $T_d$ sites are occupied by Co ions and the $O_h$ sites are shared by Ni and Co ions. Only the local environments of two $O_h$ sites are shown as the shaded polyhedrons, while the local environments of all the $T_d$ sites are shown. Right: the electronic configurations of Ni and Co ions on the $O_h$ sites and $T_d$ sites respectively. The curved arrow indicates the nearest-neighbor hopping process.

The surface morphology of films was studied by the atomic force microscopy. The electrical transport properties of the films were measured using the Van der Pauw method. The magnetic properties of the films were examined using a superconducting quantum interference device (SQUID) magnetometer. A sequence of annealing on a NCO (111)/ALO (001) film of 96 nm was carried out using a tube furnace in one atm $O_2$. For each annealing step, the film is heated for 3 hours, followed by X-ray diffraction (XRD) at room temperature and the transport measurements.

## 3. Results and discussion

### 3.1. Similar optimal growth conditions in NCO/MAO and NCO/ALO films

Previous studies have shown that both the growth temperature and the $O_2$ pressure are important factors for obtaining low resistivity in NCO epitaxial thin films [1-4]. The consensus for optimal growth temperature is about 350 ˚C [1-4,14]. We have grown NCO films in various growth temperature (300 - 500 ˚C) and $O_2$ pressure (5- 50 mTorr). The results indicate that the optimal growth temperature and O$_2$ pressure are about 360 °C and 20 mTorr respectively for both MAO and ALO substrates (**Figure** 2c, d). As shown in **Figure** 2a and b, x-ray diffraction spectra indicate no impurity phase in the NCO films of the optimal growth condition for both MAO and ALO substrates. While the NCO (111)/MAO (111) films show metallic behavior at the optimal growth condition (20 mTorr), all the NCO (111)/ALO (001) films show semiconducting behavior, even for the optimal growth condition (**Figure** 2c).

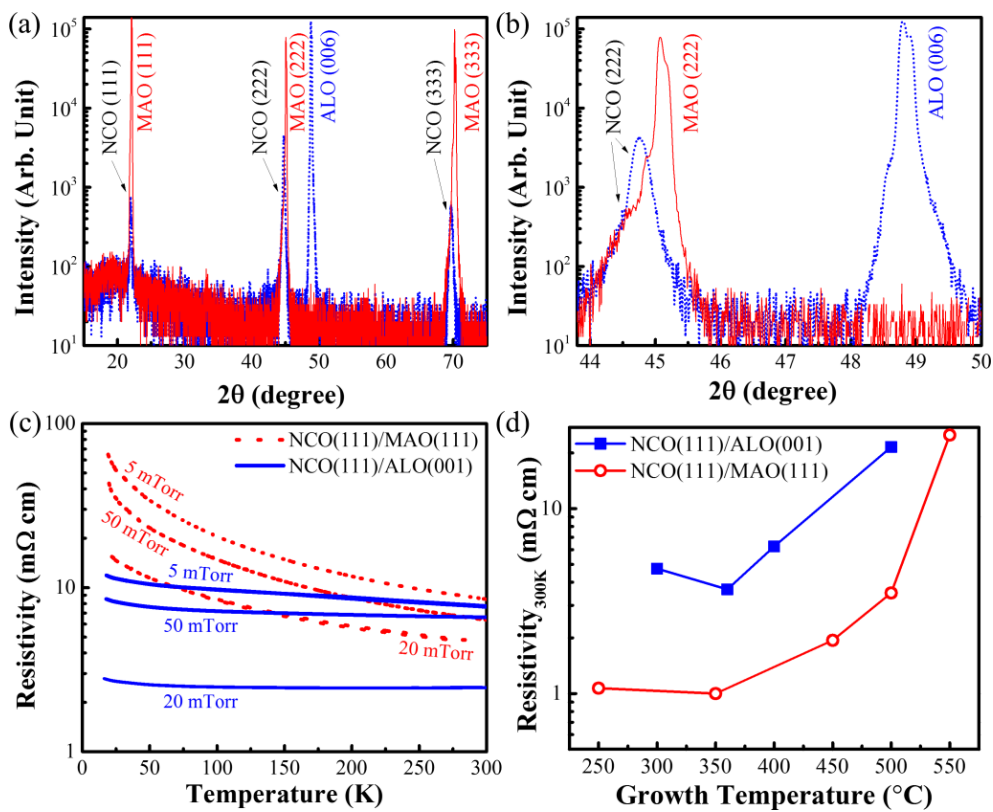

**Figure 2**. (a) θ/2θ x-ray diffraction spectra of NCO (111) films (95 nm) grown on MAO (111) and on ALO (001) substrates. (b) Closeup view of the spectra in (a) around the NCO (222) peak. (c) Temperature dependence of the resistivity of NCO (111) films at different pressure on MAO (111) and on ALO (001) substrates. (d) Resistivity at 300 K of NCO (111) films grown on ALO (001) and MAO (111) substrates.

**3.2. Hopping conduction model and effect of cation disorder**

For the NCO (111)/MAO (111) films (**Figure** 3a), the resistivity does not change greatly in the range of thickness 24 to 95 nm. In stark contrast, for the NCO (111)/ALO (001) films, the resistivity increases rapidly when the film thickness decreases; the relative increase is larger at lower temperature (**Figure** 3b).

In order to understand the mechanism of the electrical conduction in NCO (111)/ALO (001) films, we fit the temperature dependence of conductivity ($\sigma$) using the model of hopping conduction:

$$\sigma = \frac{1}{\rho} = \frac{C_0}{T}\exp(-\frac{T_{NN}}{T}) + \sigma_0 \exp\left[-\left(\frac{T_{VR}}{T}\right)^{\frac{1}{d+1}}\right]$$

where $\rho$ is the resistivity, $d$ is the dimension, $T$ is temperature, $C_0$, $\sigma_0$, $T_{NN}$ and $T_{VR}$ are the fitting parameters, of which the physical meaning will be discussed later. The first term describes the nearest-neighbor hopping and the second term describes the variable-range hopping. Previously, this model has been employed to explain the semiconducting behavior of NCO nanoplates ($d = 3$), in which the conductivity is in the range of that of the NCO (111)/ALO (001) films in **Figure** 3b [16]. As shown in **Figure** 3b, all the curves can be fit using the hopping conduction model. The results of the fitting parameters $T_{NN}$ and $T_{VR}$ are plotted in **Figure** 3 b and c. As shown in Figure 3 c, the temperature dependence of the two effects are quite different. The nearest-neighbor hopping contributes more to the conductivity at high temperature but diminishes quickly at low temperature. In contrast, the variable range hopping contributes significantly at both high and low temperature, which is an indication of the important role played by the disorder in the electrical transport. As shown in Figure 3 c, $T_{NN}$ changes slowly with the film thickness, while $T_{VR}$ changes by more than one order of magnitude for the thickness range 24 to 95 nm.

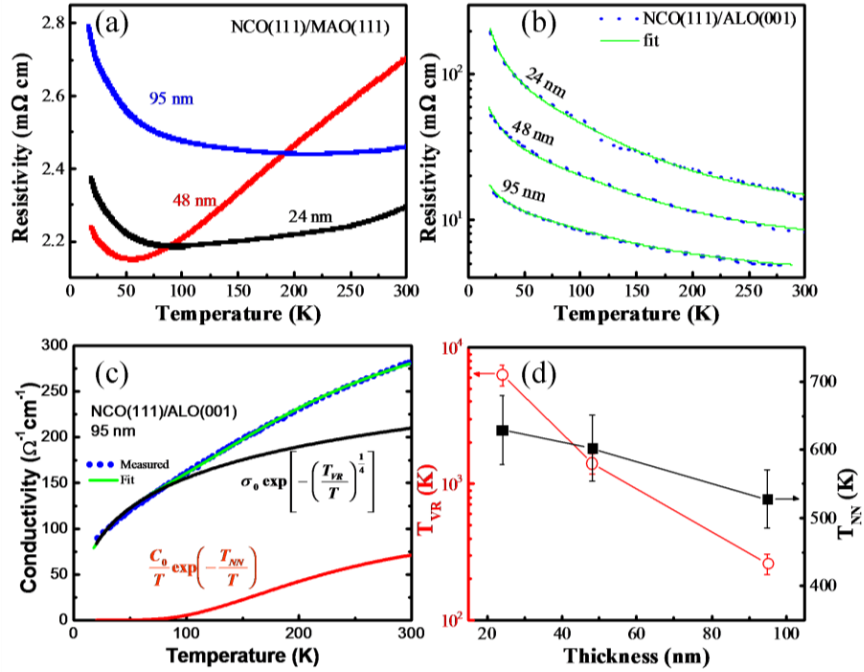

**Figure 3**. (a) Temperature dependence of NCO (111) films grown on MAO (111) of different film thickness. (b) Temperature dependence of NCO (111) films grown on ALO (001) of different film thickness. The lines are fit using the hopping conduction model (see text). (c) Fit to the conductivity of the 95 nm film in (b) using the hopping conduction model. (d) Temperature parameters found in the fit in (b). All the films were grown at the optimal conditions.

In the model of hopping conduction, $k_B T_{NN}$ is the activation energy of the nearest-neighbor hopping, where $k_B$ is the Boltzmann constant; the activation energy is found to be 54, 52, and 44 meV for the 24, 48, and 95 nm films, respectively. The previously reported nearest-neighbor hopping activation energy (66 meV) in nanoplates is also close to these values [16]. As depicted in **Figure** 1, the nearest-neighbor hopping is expected to occur between Ni ions on the $O_h$ sites. In principle, the hopping of electrons from a $Ni^{2+}$ ion to a $Ni^{3+}$ ion on the $O_h$ sites, has the same initial and final electronic configurations ($Ni^{2+}Ni^{3+}$), i.e., the same initial and final energies. On the other hand, the Ni-O bond length changes according to the valence of the Ni ion. So, the hopping of electrons changes the local Ni-O distance, creating local structural distortion (phonon); this generates an energy barrier. Therefore, the $Ni^{2+}$ to $Ni^{3+}$ hopping can be understood as a combination of electronic and vibrational excitations, or polarons. Polaron excitations have been observed in other mixed valent materials, such as $Fe_3O_4$ and $LuFe_2O_4$, with significantly larger hopping energies (about 0.16 and 0.25 eV, respectively) [29-31]. The relatively weak dependence of the activation energy on film thickness suggests that the local structure is only slightly affected by the film thickness. For the variable-range hopping, $k_B T_{VR} = \frac{24}{\pi} \frac{1}{g \xi^d}$, where $g$ is the density of state and $\xi$ is the spatial extension (size) of the localized state. The dramatic change of $T_{VR}$ suggests that the size of the localized states shrinks when the film is thinner.

Previous studies on the NCO (001)/MAO (001) films show that the cation disorder is important for metallicity, which was believed as an important reason for the low optimal growth temperature ($\approx$ 350 °C) [14,22,26,32]. In particular, the degree of spinel inversion, defined as the proportion of Ni ions on the $O_h$ site, was also found to be critical for high conductivity [14]. To investigate the effect of cation disorder on the conductivity of the NCO (111)/ALO (001) films, we studied their transport and structural properties after the post-growth annealing. As shown in **Figure** 4a, after being annealed at 500 °C, the film shows substantially reduced resistivity. We fit the resistivity using the hopping conduction model, and the results are shown in **Figure** 4b. Both the nearest-neighbor hopping activation energy $k_B T_{NN}$ and the variable hopping temperature $T_{VR}$ are reduced after the annealing at 500 ˚C. This is consistent with the previous finding that the annealing can increase the degree of spinel inversion and reduce the resistivity [14]. As shown in **Figure** 4a and b, further annealing at temperature above 600 ˚C, actually increases the resistivity, $T_{NN}$ and $T_{VR}$ again, the cause of which is revealed by the structural characterization. In **Figure** 4c, the θ/2θ x-ray diffraction spectra around the NCO (222) peak are displayed for different annealing temperature. At above 600 ˚C, the peak intensity starts to decrease; eventually the NCO (111) peak split into two peaks, indicating the decomposition of NCO into NiO and $Co_3O_4$. Although the post-growth annealing below 600 ˚C decreases the resistivity of the NCO (111)/ALO (001) films substantially, the temperature dependence of the resistivity still shows semiconducting behavior. Therefore, the degree of spinel inversion is unlikely to cause the loss of metallicity in the NCO (111)/ALO (001) films.

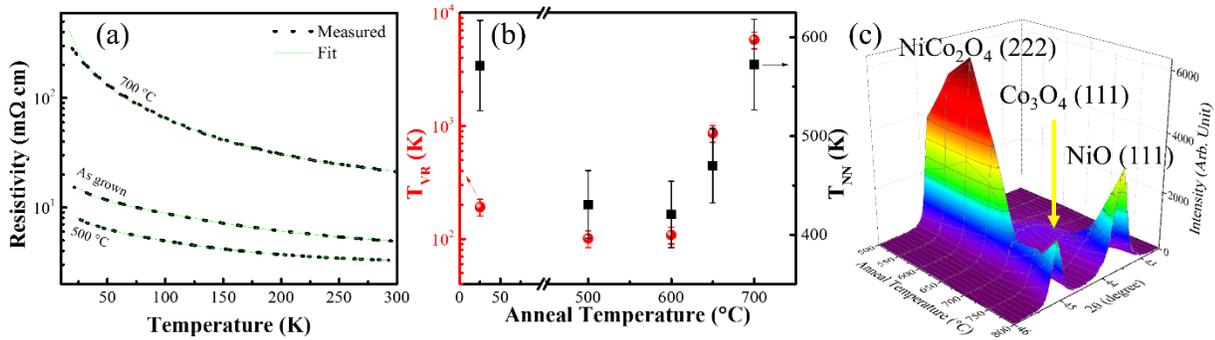

**Figure 4.** (a) Temperature dependence of resistivity of a 95 nm NCO (111)/ALO (001) film. (b) The hopping temperatures found by fitting the resistivity in (a) using the hopping conduction model. The values at 25 ˚C represent those found from the as-grown films. (c) X-ray diffraction around NCO (222) peak after post-growth annealing at different temperature.

### 3.3. Effect of valence mixing

Previous studies on epitaxial thin films indicate that the electronic structures, especially the valence mixing on Ni and Co, are critical for metallicity in NCO: In the NCO (001) /MAO (001) films, high growth temperature changes the degree of valence mixing and reduces the electrical conductivity, which is corroborated by the significant reduction of saturation magnetization [1,2]. In the NCO (001) /MgO (001) films, low $O_2$ growth environment affects the oxygen stoichiometry and reduces the electrical conductivity, which is accompanied by a large increase of saturation magnetization [3].

We measured magnetic properties of the NCO (111)/MAO (111) and NCO (111)/ALO (001) films, because they have been demonstrated to be good indicators of electronic structures, especially the valence mixing [1,3]. As shown in **Figure** 5a, the temperature dependence of magnetization of the two films follow each other rather closely. The low-temperature field dependence of magnetization shows a roughly 10% difference in the saturation magnetization and a slightly larger coercivity (**Figure** 5b). The overall differences between the magnetic properties of NCO (111)/MAO (111) and NCO (111)/ALO (001) films are modest, in comparison with the observation in the NCO (001)/STO (001) and NCO (001)/MgO (001) films [1,3]. On the other hand, the magnetoresistance (MR) of the two films, defined as $[R(H)-R(0)]/R(H)$, show dramatic differences (**Figure** 5c), where $R$ and $H$ are the resistance and magnetic field. While the NCO (111)/MAO (111) film has a small MR, in agreement with previous studies [1,2], the NCO (111)/ALO (001) film exhibits a much larger MR at low temperature. This temperature dependence and magnitude is consistent with that found in the $Fe_3O_4$ deposited on MgO substrate [33,34], which was interpreted as an effect of the anti-phase boundary in the film [35,36]. Therefore, it is unlikely that the absence of metallicity in NCO (111)/ALO (001) films is due to a significant change of electronic structures, such as the valence mixing.

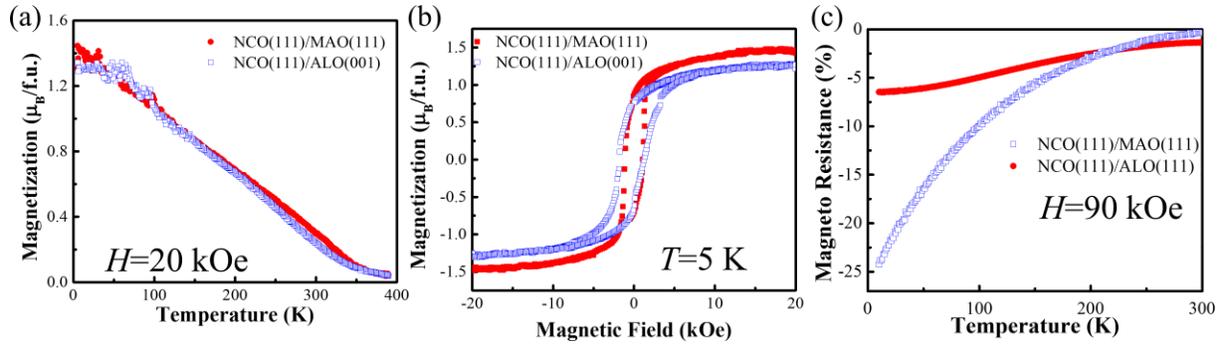

**Figure 5**. (a) Temperature dependence of magnetization of NCO (111) films (95 nm) measured in 20 kOe field on cooling. (b) Field dependence of magnetization of the NCO (111) films measure at 5 K. The films were grown at the optimal conditions. (c) Magnetoresistance defined as [R(H)-R(0)]/R(H), as a function of temperature, where $H$ is 90 kOe. The magnetic field is in the film plane. The films were grown at the optimal conditions.

## 3.4. Effect of epitaxial strain

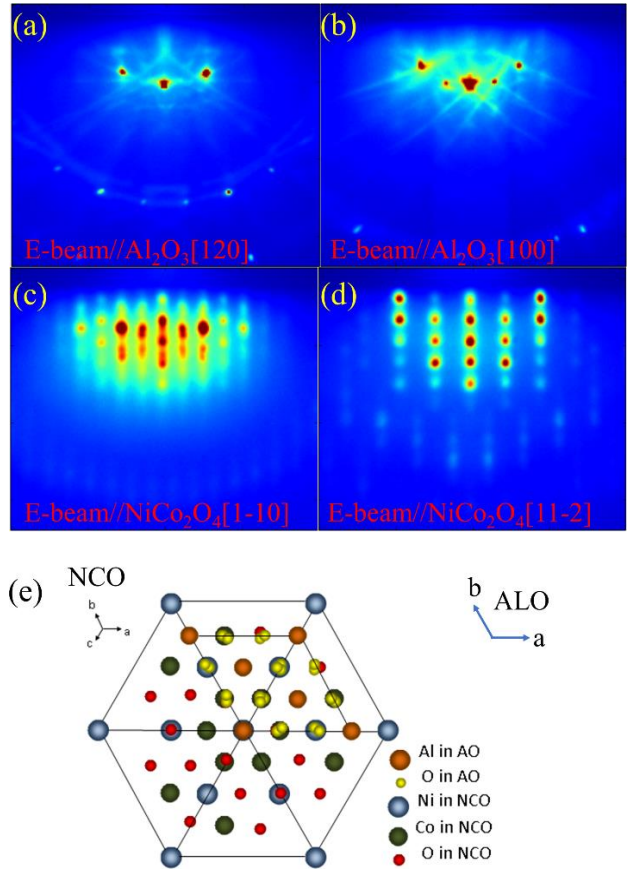

**Figure 6.** (a-d) HEED images of different surfaces with two perpendicular directions of incident electron beams relative to the substrate. In (a) and (c), the orientation of the substrate is fixed so that the electron beam is parallel to ALO [120]. In (b) and (d), the orientation of the substrate is fixed so that the electron beam is parallel to ALO [100]. The alignment between the electron beams and the NCO films lattices are also indicated. (e) Atomic model of the epitaxial relation between NCO (111) and ALO (001) planes.

Next, we investigate the correlation between the structural properties of the NCO (111)/ALO (001) films and the resistivity. We start by looking at the epitaxial relations. **Figure** 6a-d shows the diffraction pattern of the NCO (111)/ALO (001) film surface using RHEED. By comparing the diffraction pattern obtained when the electron beam is pointing in the same direction, one can obtain the in-plane epitaxial relation: ALO [120] // NCO [1-10] and ALO [100] // NCO [11-2]. The atomic arrangement of this epitaxial relation is illustrated using the models in **Figure** 6e. If we treat NCO [1-10] and NCO [01-1] as the basis of the NCO (111) plane two-dimensional unit cell, the basis of the ALO (001) plane is rotated by 30 degrees with respect to the basis of the NCO (111) plane. Similar epitaxial relation has been observed at the ALO (001) / $Fe_3O_4$ (111) interface, which was explained in terms of the matching of oxygen sublattice [37,38].

One may estimate the possible epitaxial strain from **Figure** 6e using the small mismatch of the oxygen sublattice in the ALO (001) plane and that of the NCO (111) plane; the result is a 4% in-plane compressive strain for the NCO (111) films. On the other hand, this strain appears to be mostly relaxed in the NCO (111)/ALO (001) films we have grown. As shown in **Figure** 7a, the *d*-spacing of the NCO (111) plane was measured using the θ/2θ x-ray diffraction and the out-of-plane lattice constant were calculated for NCO films grown at various conditions. The out-of-plane lattice constant of the NCO (111) films grown on ALO (001) is only 0.4% larger than that of the bulk NCO, which is incompatible to the speculated 4% in-plane compressive strain. Therefore, the strain in the NCO (111)/ALO (001) films is mostly relaxed. In contrast, for the NCO (111)/MAO (111) films, the out-of-plane lattice constant is about 1% larger than that of bulk NCO, which agrees with that in the NCO (001)/MAO (001) films [1], indicating a small unrelaxed compressive strain up to at least 95 nm of film thickness. In addition, according to **Figure** 7a, the out-of-plane lattice constant of NCO (111) film grown on ALO (001) appears to be not sensitive to the growth temperature and the film thickness. Therefore, it is unlikely that the absence of metallicity of NCO (111)/ALO (001) films is due to the epitaxial strain.

### 3.5. Effect of nanostructural disorder

According to **Figure** 6e, the size of the in-plane unit cell of the NCO (111) plane and that of the ALO (001) plane do not have a one-to-one matching relation. This large difference in the size of the unit cells will generate structurally incompatible interfaces between crystallites nucleated at random positions. These interfaces, also called anti-phase boundaries, are expected to complicate the nanostructure, which was investigated here in the NCO (111)/ALO (001) films by the electron and x-ray diffractions. As shown in **Figure** 6c and d, the RHEED images of the NCO (111)/ALO (001) films show typical patterns for quasi two-dimensional (2D) morphology, in that the vertical streaks and the arch-shaped arrangement indicate 2D reflection while the array of diffraction spots indicates formation of small islands. Further information on the nanostructure can be inferred from the structural correlation of the NCO films extracted from the x-ray diffractions. The out-of-plane structural correlation length can be estimated from the width of the $\theta/2\theta$ x-ray diffraction peaks, and the result is shown in **Figure** 7b. As the film thickness increases, the structural correlation length also increases; the values are always smaller than the film thickness. The in-plane structural correlation length can be found from the x-ray diffraction of the films (**Figure** 7c): the rocking curve of the 48 nm and 95 nm NCO (111)/ALO (001) films show a narrower peak standing on a broader peak, indicating two types of in-plane structural correlation length. While the longer in-plane correlation length (extracted from the narrower peak) increases with the film thickness, the shorter correlation length (extracted from the broader peak) remain relatively constant (about 12 nm) for the 48 and 95 nm films. Surface morphology of the NCO (111)/ALO (001) films was measured using the atomic force microscopy (**Figure** 7d). Small crystallites of about 10 nm are observed at the film surface, which is consistent with the shorter structural correlation length found by the x-ray diffraction. Hence, the structural correlation length, both in-plane and out-of-plane increases with the film thickness.

The nanostructural disorder caused by the anti-phase boundaries may account for all the above observations on the NCO (111)/ALO (001) films, as discussed below.

Under the same optimal growth conditions, the electronic structures and the cation distribution within every nano-sized crystallite of the NCO (111)/ALO (001) films is likely to be similar to those of the NCO (111)/MAO (111) films. Therefore, the magnitude and temperature dependence of the magnetization of the NCO (111)/ALO (001) films, which reflects the local properties of the films, are expected to be similar to those of the NCO (111)/MAO (111) films; this agrees with the observation in **Figure** 5a and b. On the other hand, for the electrical conduction, the interfaces (anti-phase boundaries) between crystallites play extremely important roles. Due to these anti-phase boundaries, the electrons get localized and adopt the hopping mechanism for conduction; the spatial extension of the localized states in the variable-range hopping is then related to the size of the crystallite. According to the analysis in **Figure** 3b and c, $T_{VR}$ decreases with film thickness, indicating that the spatial extension of the localized states $\xi$ increases with the film thickness; this is consistent with the observation (**Figure** 7b) that the correlation length of the film, which is related to the size of the crystallites, increases with the film thickness.

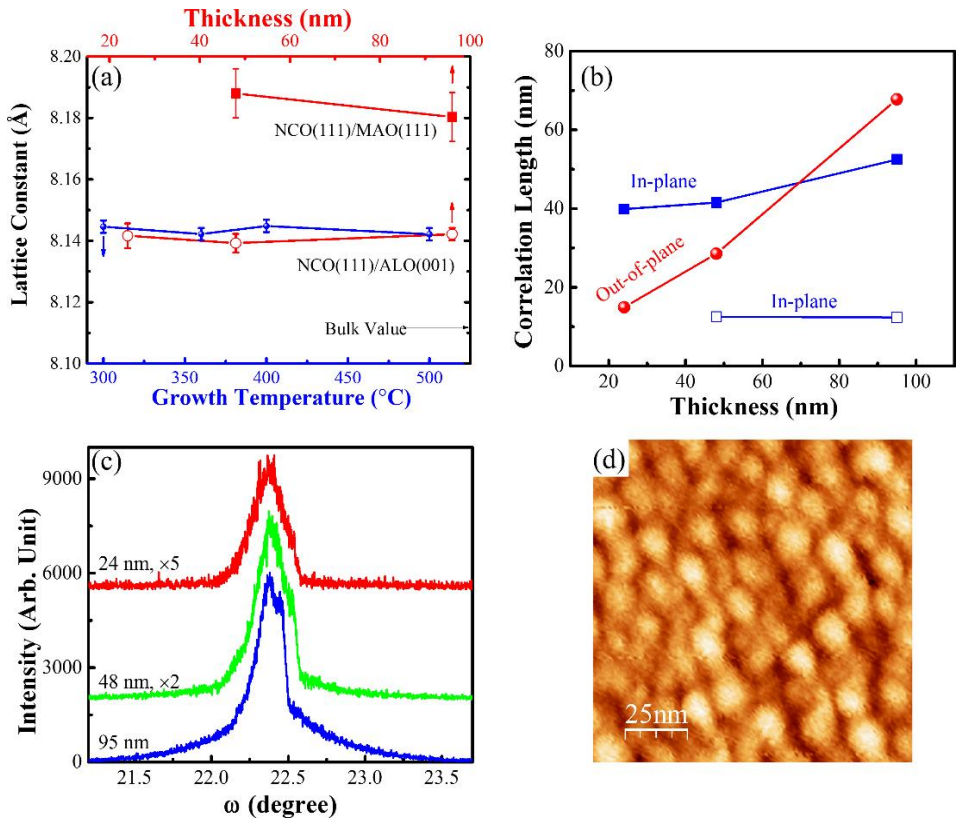

**Figure 7.** (a) Out-of-plane lattice constant of the NCO (111) films for different substrates and at different growth conditions. (b) The in-plane and out-of-plane structural correlation length of the NCO (111)/ALO (001) films as a function of film thickness. (c) Rocking curve of NCO (111)/ALO (001) films of different film thickness. (d) Atomic force microscopy image of a NCO (111)/ALO (001) film (95 nm). The films in (b)-(d) were grown at the optimal conditions.

Because the anti-phase boundaries originate from the structural mismatch between the film and the substrates and the random location of the nucleation during the film growth, post-growth annealing is unlikely to remove the anti-phase boundaries. Therefore, the spatial extension of the localized states is not expected to change significantly during the annealing; this is corroborated by the observation of modest change of $T_{VR}$ after the annealing up to 600 °C. In contrast, the crystallinity within every crystallite could be improved by the annealing, as indicated by the obvious change of $T_{NN}$ after the annealing up to 600 °C, because high crystallinity is expected to reduce the barrier of the nearest-neighbor hopping.

The electrical conduction between crystallites may also depend on the relative spin alignment between the two sides of the interfaces. In other words, the crystallite/anti-phase boundary/crystallite may behave like a spin valve. Since the initial and final states of the hopping are similar states at different sites, a parallel spin alignment is favored for lower resistance. Thus, negative magnetoresistance is expected, which is consistent with the observation in **Figure** 5c. In this case, more interface causes larger magnetoresistance, which is why the magnetoresistance is much larger in the NCO (111)/ALO (001) films than that in the NCO (111)/MAO (001) films.

**4. Conclusions**

To study the metallicity in NCO, we have compared the epitaxial NCO (111) films grown on MAO (111) and ALO (001) substrates. Despite the same optimal growth conditions, the NCO (111)/MAO (111) films are metallic while the NCO (111)/ALO (001) films are semiconducting. The magnetic properties and the effect of post-growth annealing suggest that the known mechanism for absence of metallicity in NCO, such as the deviation from the optimal valence mixing and the optimal cation occupancy, are not the origin of semiconductivity in NCO (111)/ALO (001) films. On the other hand, the presence of the anti-phase boundaries, which originate from the mismatch between the crystal structures of NCO and ALO and the random nucleation during the film growth, may explain all the observations in the NCO (111)/ALO (001) films, including the thickness dependence of the resistivity, the effect of post-growth annealing, the similarity between the magnetization of the NCO (111)/MAO (111) films and that of the NCO (111)/ALO (001) films, the sign and large magnitude of the magnetoresistance in NCO (111)/ALO (001) films. Therefore, we propose that the nanostructural disorder caused by the anti-phase boundaries between the crystallites in the NCO (111)/ALO (001) films, is the main factor for the absence of metallicity. These findings shed important light on the absence of metallic behavior of NCO in various forms, especially in polycrystalline samples. The large magnetoresistance in the NCO/ALO films could be exploited for potential spintronic applications.

**Acknowledgments**

This project was primarily supported by the National Science Foundation (NSF), DMR under Award DMR-1454618 and by the Nebraska Center for Energy Sciences Research. C. M. Zhen would like to acknowledge the support provided by the Key Project of Natural Science of Hebei Higher Education under Grant No. ZD2017045. The research was performed in part in the Nebraska Nanoscale Facility: National Nanotechnology Coordinated Infrastructure and the Nebraska Center for Materials and Nanoscience, which are supported by the National Science Foundation under Award ECCS: 1542182, and the Nebraska Research Initiative.